# Towards Replacing Resistance Thermometry with Photonic Thermometry

Nikolai Klimov[1,2], Thomas Purdy[3] and Zeeshan Ahmed[2*]

[1]Joint Quantum Institute, University of Maryland, College Park, MD 20742
[2]Thermodynamic Metrology Group, Sensor Science Division, Physical Measurement Laboratory, National Institute of Standards and Technology, Gaithersburg, MD 20899
[3]Quantum Optics Group, Quantum Measurement Division, Physical Measurement Laboratory, National Institute of Standards and Technology, Gaithersburg, MD 20899
[*]zeeshan.ahmed@nist.gov

**Abstract:** Resistance thermometry provides a time-tested method for taking temperature measurements that has been painstakingly developed over the last century. However, fundamental limits to resistance-based approaches along with a desire to reduce the cost of sensor ownership and increase sensor stability has produced considerable interest in developing photonic temperature sensors. Here we demonstrate that silicon photonic crystal cavity-based thermometers can measure temperature with uncertainities of 175 mK ($k = 1$), where uncertainties are dominated by ageing effects originating from the hysteresis in the device packaging materials. Our results, a ≈ 4-fold improvement over recent developments, clearly demonstate the rapid progress of silicon photonic sensors in replacing legacy devices.

## 1. Introduction

Temperature measurements play a central role in all aspects of modern life ranging from process control in manufacturing[1], physiological monitoring[2,3] in medicine, to environmental engineering control in buildings[4] and automobiles[5]. Despite the ubiquity of thermometers, the underlying technology, resistance measurement of a thin metal film or wire, has only undergone incremental improvements over the last century[6,7]. Though resistance thermometers can routinely measure temperature with uncertainties as low as 10 mK, they are sensitive to environmental variables such as humidity, chemical oxidation and mechanical shock which causes the resistance to drift over time, requiring frequent off-line, expensive, and time consuming calibrations[6]. In recent years, there has been considerable interest in developing photonic devices as an alternative to resistance thermometers as a means of overcoming the shortfalls of resistance thermometry. Optical fiber and silicon photonic based technologies have received considerable interest as they have the potential to provide greater temperature sensitivity while being robust against mechanical shock and electromagnetic interference. Furthermore, the low weight, small form factor photonic devices may be multiplexed to provide low-cost sensing solutions[8-10].

Photonic temperature sensors exploit temperature dependent changes in a material's properties – typically, a combination of thermo-optic effect and thermal expansion [11,12]. For example, fiber Bragg gratings (FBG), exhibit temperature dependent shifts in resonant Bragg wavelength of ≈ 10 pm/K [8,13,14] which can be utilized to measure temperature over the range of 233 K to 293 K with uncertainties of 500 mK ($k = 2$) when humidity and strain effects are minimized [14]. The impact of humidity on silicon photonic devices' performance is significantly reduced by depositing a passivating silicon dioxide layer on top of the silicon device [10]. Silicon Bragg thermometers have been demonstrated to measure temperature with uncertainties of 1.25 K ($k = 2$) over the range of 278 K to 433 K [15]. The measurement uncertainty in silicon Bragg devices is dominated by the uncertainty in peak center measurement which could be significantly reduced by fabricating high quality-factor (Q-factor) devices such as photonic crystal cavity or ring resonators.

Numerous researchers have reported on silicon ring resonator and photonic crystal cavity (PhCC) based temperature sensors that were probed using fiber-to-chip evanescent coupling cumulatively demonstrating the superior temperature sensitivity, noise floor and temporal response of silicon photonic temperature sensors [10,16-20]. Recently, we undertook a systematic survey of ring resonator parameter space that aimed to optimize the device performance while achieving consistent results.[21,22] Our results suggest that consistently high performance temperature sensors are obtained from the zone of stability (waveguide width > 600 nm, air gap ≈ 130 nm and ring radius > 10 μm) such that quality factors are consistently $10^4$ and the temperature sensitivity is consistently in the 70 pm/K to 85 pm/K range.[22] For evanescently coupled devices, over the temperature range of 293 K to 418 K, the fit residual varies between 170 mK to 30 mK. Although comparison of the same device fabricated across different chips in the same batch reveals significant variation in temperature response, our results suggest that with better process control it is possible to achieve device interchangeability over a 200 mK tolerance band i.e. devices of the same design parameter, using nominal calibration coefficients will provide temperature readings that are within 200 mK of each other. Similar results can be reasonably expected of other resonant Si devices such as PhCCs.

Here we build upon our recent progress and characterize the temperature response of a packaged PhCC device over an extended temperature range. Our results indicate that these silicon photonic devices enable measurements of temperature with uncertainties of 175 mK ($k = 1$). The long-term stability of the thermometer is limited by hysteresis likely due to the epoxy used in pigtailing of fiber array to the chip. Nevertheless, this result represents a ≈ 4-fold improvement over the Si Bragg waveguide thermometer[15].

## 2. Experimental:

The photonic chip with integrated temperature sensors was fabricated at the National Institute of Standards and Technology (NIST), Center for Nanoscience and Technology (CNST) using the CMOS process line on a silicon-on-insulator (SOI) wafer with a 220 nm thick layer of crystalline silicon on top of a 3 μm thick buried oxide layer that isolates the optical mode and prevents light leakage into the substrate. The device is initially patterned via electron beam lithography followed by an inductive coupled plasma reactive ion etch (ICP RIE) of 220 nm-thick silicon layer of the SOI substrate. After silicon etch an 800 nm-thick passivating layer of silicon dioxide was deposited via plasma enhanced chemical vapor deposition (PECVD).

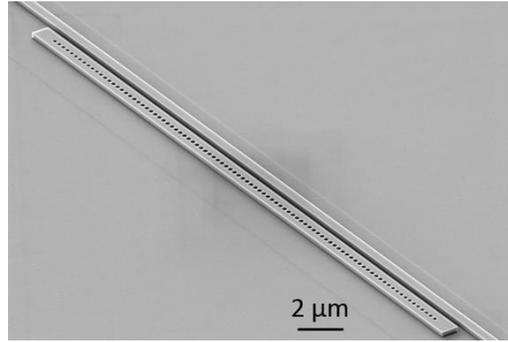

Fig 1: SEM image of a silicon nanobeam photonic crystal cavity (Si PhCC) device.

The photonic thermometer described in this work is a Si PhCC device (Fig. 1), operating in the telecom frequency range. The PhCC, similar to a macroscopic Fabry-Perot cavity, is sensitive to changes in effective length. Linear expansion due to temperature or strain and local changes in refractive index due to temperature (thermo-optic effective), strain and/or environment (e.g. humidity or gas pressure) lead to mode frequency shift. For the case of a temperature sensor, the goal is to minimize the impact of other variables while leveraging the thermo-optic coefficient of silicon to realize highly sensitive thermometers. Silicon with a thermo-optic coefficient that is ≈100X larger than its linear expansion coefficient, is one of the best materials to build a photonic thermometer. The sensor consists of a silicon waveguide ($w = 800$ nm; $h = 220$ nm) that has a cavity at its center. The design of the cavity follows a deterministic approach of Quan *et. al* [23]. Two symmetrical Bragg mirrors made of one-dimensional array of holes etched in the nano-waveguide are placed on the opposite sides from the cavity. The diameters of the holes in the Bragg mirrors are monotonically tapered from 170 nm, at the edge, to 200 nm, at the center, to achieve a Gaussian field profile within the cavity, which minimizes radiation losses and maximizes the $Q$ of the cavity [23]. The light was coupled into the cavity via an evanescent coupling from an adjacent 510 nm-wide bus waveguide placed within 250 nm of the PhCC (Fig. 1). Focusing grating couplers were utilized as a means for efficient coupling of light from an optical fiber in/out of the photonic device with coupling losses on the order of -4 dB per coupler.

The photonic chip with integrated PhCC temperature sensors was pigtailed using v-groove fiber array and UV curable adhesive. Following this step the photonic chip was lowed into a glass tube (with a diameter of approximately 12 mm). A small amount of dry magnesium oxide (MgO) is added to the glass tube to improve heat-exchange between the chip and the tube walls. Lastly, the tube was back filled with argon gas and sealed with epoxy.

In our experiments, a tunable extended cavity diode laser (New Focus TLB-6700)[1] was used to probe the photonic devices. A small amount of laser power was immediately picked up from the laser output for wavelength monitoring (HighFinesse WS/7) while the rest, after passing through the photonic device via grating couplers was detected by a large sensing-area power meter (Newport, model 1936-R). The assembled photonic thermometer was placed in a cylindrical aluminum block (25 mm diameter, 170 mm length). The cylinder has two 150 mm deep blind holes for accommodating a calibrated thin film platinum resistance thermometer (PRT) and the sealed in a glass tube photonic thermometer, respectively. The metal block along with the sealed in glass tube photonic thermometer was placed inside the dry-well calibrator (Fluke 9170). The dry-well calibrator's temperature was controlled by an automated LabVIEW program, which thermally cycled the bath between measured temperatures. The program also added a 30 min settling delay for temperature equilibration between the bath and the photonic thermometer. Five consecutive scans were recorded at each temperature and the photonic sensor was thermally cycled three times and once in the follow-up cycles 800 hours later. The recorded data was fitted using a piecewise polynomial fitting routine to extract peak center, peak height and peak width as a function of temperature. The peak temperature (343 K) is kept well below the glass transition temperature of the UV-curable epoxy (404 K) to ensure thermo-mechanical stability of the packaged device.

**3. Results and Discussion:**

Figure 2A shows transmission spectra of PhCC measured in the wavelength range between 1520 nm and 1570 nm. In the measured wavelength range the PhCC has four different modes with Q-factors of $\approx 10^4$ and a free spectral range of $\approx 9.4$ nm. All four modes show significant redshift at higher temperatures. At telecom frequency range, resonator-based Si photonic devices have been demonstrated to exhibit self-heating due to a two-photon absorption[24] that leads to a laser power dependent redshift of the resonance wavelength. As shown in Figs. 2C-D, our photonic thermometer exhibits similar self-heating behavior with the optical bistability evident at input powers of $> 40$ µW. The self-heating induced redshift increases linearly with the input laser power at a rate of $\partial(\delta\lambda)/\partial P \approx 3.65$ pm/µW or $\approx 48.5$ mK/µW (Fig. 2E) and is independent of background gas pressure (Fig C-E). Similar to resistance thermometry, the effect of self-heating in a photonic thermometer gives rise to a fixed offset error that can be mitigated by operating the optical thermometer at the calibration power or by making a power dependent correction. The former approach is a common practice in resistance thermometry, where the calibration of resistance thermometers and their usage are performed at the same electrical currents to mitigate the current-induced self-heating. We estimate the noise floor using a side-of-fringe measurement (see ref 10). Figure 2F shows the Allan deviation plot of the PhCC thermometer (red curve) which shows a noise floor of 70 µK at 0.1 s integration time and 2 kHz data acquisition rate. Dash-dotted line represents a long-term drift in the system of 4 mK/hr.

---

[1] Disclaimer: Certain equipment or materials are identified in this paper in order to specify the experimental procedure adequately. Such identification is not intended to imply endorsement by the National Institute of Standards and Technology, nor is it intended to imply that the materials or equipment identified are necessarily the best available.

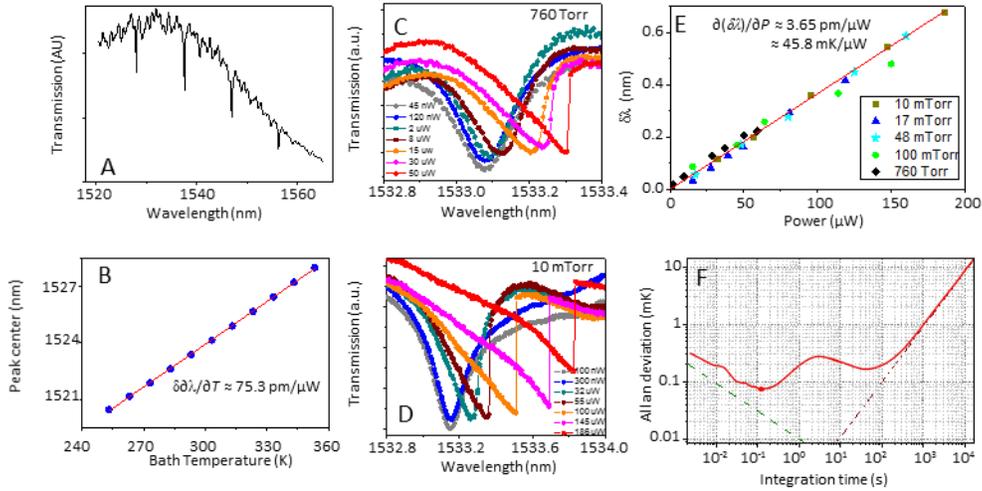

Fig 2: (A) Transmission spectra of PhCC at 353.15 K. (B) Redshift of the resonance wavelength of the fundamental cavity mode with increasing temperature (blue dots) and a corresponding linear fit (red curve) with 75.3 pm/K slope. (C, D) Dependence of the transmission spectra on the input laser power measured at (C) 760 Torr and (D) 10 mTorr ambient gas pressure. (E) Self-heating of the fundamental cavity mode: the redshift, $\delta\lambda$, of the resonance wavelength vs. input laser power measured at different ambient gas pressures. Red curve is a linear fit to the data. (F) Allan deviation plot of the measured temperature (red curve) shows that at a noise floor of 70 µK (red dot) at 0.1 s integration time and 2 kHz acquisition rate. Dashed green line represents detector noise floor. Dash-dotted red curve shows 4 mK/hr resonance wavelength drift.

To test the performance of the packaged PhCC thermometer, we thermally cycled in a dry well temperature calibrator. The cycling was performed four times. Each cycle/run consisted of two ramps: ramp up (from 293.15 K to 343.15 K) and a ramp down (from 343.15 K to 293.15 K), each performed in 5 K steps. For easier reference we denote each cycle as $Run_{i,j}$, where the first subscript represents the cycle number ($i$ = 1, 2, 3, 4) while the second subscript shows the direction of the ramp ($j$ = "up", "down"). After every 5 K incremental step the calibrator's temperature, $T_{bath}$, was stabilized and measured by a calibrated thin film PRT.

In order, to minimize the impact of self-heating on the temperature measurement, while maintaining a good signal-to-noise ratio, thermal cycling experiments were carried out at 200 nW input laser power. At this power level, a 1 % - fluctuation in incident laser power leads to ≈ 0.1 mK measurement uncertainty ($k$ ≈ 1) attributed to uncontrolled self-heating fluctuations. To further decrease the measurement uncertainty, five consecutive transmission spectra were taken around the fundamental mode of the cavity at each temperature. For each ramp the measured data set, ($\lambda_{PhCC}$, $T_{bath}$), shows a nearly linear dependence of ≈ 75.3 pm/K (Fig. 2B). The temperature response is best modeled as a second order polynomial. Figure 3A shows the six fit residuals (*FitRes*) from quadratic fits to each temperature ramp. The averaged standard deviation of the fit residuals from a quadratic fit is 43 mK. The residuals show the greatest variability in the very first temperature ramp, $Run_{1,up}$, with deviations peaking at 308.15 K, resulting in a relatively poor fit. The standard deviation for the fit residuals is 86 mK. In all subsequent temperature ramps (up or down), the fit residuals are closely packed with standard deviations of each ramp distributed between 30 mK and 45 mK (Fig. 3A). Our results indicate that short term hysteresis is driven by initial changes encountered in the first ramp only (Fig 2B)[2]. We explored long-term hysteresis by examining the temperature response of the device 800 hours later ($Run_4$). As shown in Fig 3C the calculated residuals for the third and fourth ramps show similar profiles; standard deviations of fit residuals for both cycles are nominally the same, 40 mK ($k = 1$). However, the residuals for the fourth run show a 170 mK offset from $Run_3$. .

---

[2] The plot in Fig 2B is calculated by taking the fit parameter or $Run_{1,up}$ and calculating the deviations of measured and "predicted" values in all subsequent runs.

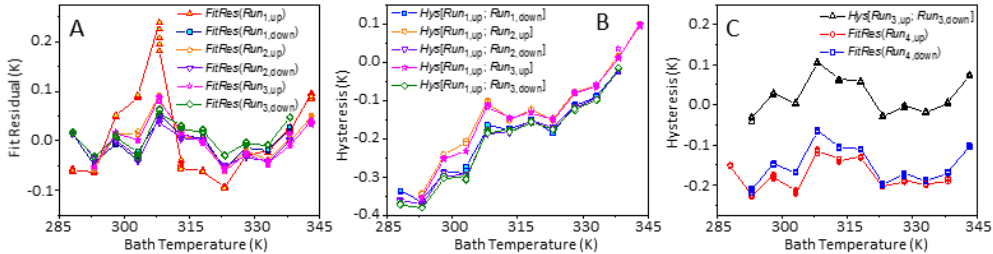

Fig 3: (A) Fit residuals, *FitRes*, for the first six thermal ramps of PhCC thermometer ($\lambda_{PhCC}$ vs. $T_{bath}$). Residual of the very first heating ramp, *FitRes*($Run_{1,up}$), exhibits significant deviations from the rest. (B) Hysteresis between the first ramp and all subsequent ramps, *Hys*[$Run_{1,up}$, * ], reveals significant ageing during the initial heating ramp, $Run_{1,up}$. (C) 800 hour later the fit residuals $Run_{4,up}$ and $Run_{4,down}$ show similar behavior as observed in "post-aged cycles of (A), $Run_2$ and $Run_3$, with no significant thermal hysteresis during the cycle. However, we do observe significant ageing between $Run_4$ and the previous three runs.

Overall, we find that long term drift in packaged devices dominates the uncertainty of PhCC (Table 1), limiting the combined expanded uncertainty to 175 mK ($k = 1$). The fit residuals from a quadratic fit suggest that a device with Q ≈ $10^4$ has a measurement accuracy of 42 mK. However, we note that the residuals show a reproducible higher order structure. The use of a higher order polynomial or a linear piece-wise fitting model could further decrease the measurement uncertainty contributed by the fit residual. Additionally, a higher Q device could be employed to improve the accuracy of peak center determination.

**Table 1: Uncertainty components in PhCC Thermometers**

| Component | Uncertainty, $k = 1$ (mK) |
|---|---|
| Temperature (PRT) | 10 |
| Wavelength | 0.53 |
| Fit Residual | 42 |
| Ageing Effect/Drift | 170 |
| self-heating | 0.1 |
| Combined Uncertainty ($k = 1$) | 175 |

We note that passivated Si sensors have been demonstrated to be exceptionally stable over months-long time-period. Furthermore, in our previous experiments where photonic temperature sensors were interrogated using free space coupling we did not observe any significant hysteresis over a week-long observation period [15]. We hypothesize that the observed thermal hysteresis/ageing is likely due to residual strain imparted by the epoxy used in packaging of the pigtailed photonic thermometer. Use of low thermal expansion epoxy with low hygroscopicity, better sealing of photonic thermometers from ambient air and sensor design evolution to minimize interactions with epoxy regions could potentially remove this significant source of uncertainty and provide a thermometer that is exceptionally stable over thousands of hours. Experiments with a new design incorporating some of these proposed changes are underway.

## 4. Summary:

We have demonstrated that PhCCs can be used for temperature sensing with uncertainties of only 175 mK ($k = 1$), a nearly 4-fold improvement over our previous designs [15]. The measurement uncertainty is dominated by thermal hysteresis and ageing effects that result in significant long term drift. The packaging's ageing effects could be mitigated with some minor sensor design and packaging changes.

**Acknowledgements**

The authors acknowledge the NIST/CNST NanoFab facility for providing opportunity to fabricate silicon photonic temperature sensors. The authors also thank Wyatt Miller and Dawn Cross for assistance in setting up the experiments and William Guthrie for helpful discussions.